\documentclass[12pt]{amsart}

\usepackage{amsmath,amsthm,amscd,euscript}
\def \r{\mathbb R}

\def \z{\mathbb Z}

\newtheorem{theorem}{Theorem}[section]

\newtheorem{statement}[theorem]{Statement}

\theoremstyle{remark}
\newtheorem{remark}[theorem]{Remark}
\newtheorem{definition}[theorem]{Definition}

\newtheorem{problem}{Problem}

\title{M\"obius energy of graphs.}
\author{Oleg~Karpenkov}
\date{14 September 2005}
\thanks{Partially supported
by RFBR grant SS-1972.2003.1 and by RFBR grant 05-01-01012a.}

\keywords{Energies of graphs, variational principles, M\"obius
transformation.}

\email[Oleg Karpenkov]{karpenk@mccme.ru}

\address{Poncelet Laboratory (UMI 2615 of CNRS and Independent University
of Moscow)}

\begin{document}
\input epsf

\maketitle

\sloppy \normalsize

\tableofcontents

\section*{Introduction.}

The study of knot energies was initiated by the work of Moffatt
(1969)~\cite{Moff1}, and was developed by him in ~\cite{Moff2}
following Arnold's work~\cite{Arn1}. The first discrete energy of
knots were produced by W.~Fukuhara in~1988, for the details see
his work~\cite{pol}. M\"obius energy was discovered by
J.~O'Hara~\cite{O-H1} in 1991. Further investigations of M\"obius
energy properties were made by M.~H.~Freedman, Z.~-H.~He, and
Z.~Wang in~\cite{Freed}. Particularly, the authors introduced
variational principles for M\"obius energy and found some upper
estimates for the minimal possible energy of knots with the given
crossing number in their work. Conformal properties of M\"obius
energy allow us to calculate explicitly some critical values for
toric knots, see the work~\cite{Kim}. The following articles were
dedicated to general theory of knot energies:~\cite{Abr},
\cite{O-H2}, \cite{O-H3}, \cite{EKar1}, \cite{EKar2}, \cite{Sul}
etc. Recently A.~Bobenko in~\cite{AIB} introduced M\"obius energy
for simplicial surfaces. A good overview of properties for knot
energies, the techniques of the approximation constructions of
extremal knots, and some generalizations of energies can be found
in the book of J.~O'Hara~\cite{O-H4}.

In the present paper we introduce {\it M\"obius energy for the
embedded graphs}. This energy is invariant under M\"obius
transformations. This paper is organized as follows. In Section~1
we give the definition of M\"obius energy. Further in Section~2
we formulate the main properties of this energy and outline the
main ideas of their proofs. In the last section we study critical
configurations for the angles at vertices of degree less than
five. We conclude the paper with a few words about the techniques
of construction of symmetric toric embedded graphs with critical
values of M\"obius energy.

The author is grateful to A.~Sossinski for constant attention to
this work and useful remarks and discussions.

\section{Definition of M\"obius energy for the graphs.}

We will start with the definition of M\"obius energy for knots in
a conformal form proposed by P.~Doyle and O.~Schramm (see, for
example,~\cite{O-H4}, page~39). In the present paper by oriented
{\it knot} we mean the $C^{2}$-smooth embedding of the circle
$S^1=\r / (2\pi \z)$ in $\r^3$. Let $\tau:S^1 \longrightarrow \r
^3$ be an oriented knot. Denote by
$C(\tau(t_1),\tau(t_1),\tau(t_2))$ the circle (or the line)
tangent to the knot at the point $\tau(t_1)$ and containing the
point $\tau(t_2)$. We orient this circle such that the obtained
orientation coincides with the knot orientation at the tangency
point $\tau(t_1)$. Denote the angle between the oriented circles
$C(\tau(t_1),\tau(t_1),\tau(t_2))$ and
$C(\tau(t_2),\tau(t_2),\tau(t_1))$ by $\theta_{\gamma}(t_1,t_2)$.
By definition the angle $\theta_{\gamma}$ is from the segment
$[0,\pi]$. By $|*|$ we denote the absolute value of the vector in
$\r^3$.

By {\it M\"obius energy} of the knot $\tau$ we mean the following
value:
$$
M(\tau)=\iint \limits_{S^1 \times S^1}\left(
\frac{\dot{\tau}(t_1)\dot{\tau}(t_2)}{|\tau(t_1)-\tau(t_2)|^2}
-\cos\theta_{\tau}(t_1,t_2)\frac{\dot{\tau}(t_1)\dot{\tau}(t_2)}{|\tau(t_1)-\tau(t_2)|^2}\right)
dt_1dt_2.
$$
M\"obius energy is well-defined, positive, does not depend on
orientation and para\-met\-ri\-za\-tion of the knot. If we
$C^2$-smoothly perturb the knot then M\"obius energy changes in
the continuous way. The energy is invariant under the group of
M\"obius transformations (i.e. the group of transformations
in~$\r^3$ generated by all inversions), see~\cite{Freed}. The
minuend of the integrand is called the {\it principal term}, the
subtrahend is called the {\it normalization term}.

Suppose now we have some graph $G$ with edges $e_i$, $i{=}1,
\ldots , n$, and vertices $v_k$, $k{=}1, \ldots , m$. For the
simplicity we suppose that the graph $G$ does not contain loops
and multiple edges (note that the construction below can be easily
generalized to the arbitrary graph). So let  $\gamma:G
\longrightarrow \r ^3$ be an embedding or an immersion that is
$C^2$-smooth on open edges and such that at any vertex the
one-sided first and second derivatives are well-defined and
continuous. Suppose also that for any couple of edges $e_i$ and
$e_j$ adjacent to the same vertex the angle $\alpha_{ij}$ between
the vectors of the corresponding one-sided first derivatives at
this vertex is non-zero. Denote the set of all angles
$\alpha_{ij}$ for all vertices of the graph $G$ by $\alpha$. Such
embedding or immersion is called an {\it$\alpha$-embedding} or an
{\it$\alpha$-immersion}. Note that for any fixed set $\alpha$ of
angles, $C^2$-topology on the space of all $\alpha$-embeddings
(or all $\alpha$-immersions) is defined in the natural way.

First, we define M\"obius energy for edges and for couples of
edges.

{\bf 1).} M\"obius energy for some edge $e_i$ is calculated
similar to the case of M\"obius energy for knots:
$$
M(\gamma;e_i,e_i)=\iint \limits_{e_i \times e_i}\left(
\frac{\dot{\gamma}(t_1)\dot{\gamma}(t_2)}{|\gamma(t_1)-\gamma(t_2)|^2}
-\cos\theta_{\gamma}(t_1,t_2)\frac{\dot{\gamma}(t_1)\dot{\gamma}(t_2)}{|\gamma(t_1)-\gamma(t_2)|^2}\right)
dt_1dt_2.
$$

{\bf 2).} Let edges $e_i$ and $e_j$ do not have any common
vertex. Orient them in an arbitrary way and define
$$
\begin{array}{l}
 M(\gamma;e_i,e_j)=\\
\displaystyle
 \iint \limits_{e_i \times e_j}\left(
\frac{\dot{\gamma}(t_1)\dot{\gamma}(t_2)}{|\gamma(t_1)-\gamma(t_2)|^2}
-\bigl( \cos\theta_{\gamma}(t_1,t_2)+\cos(\pi {-}
\theta_{\gamma}(t_1,t_2)) \bigr)
\frac{\dot{\gamma}(t_1)\dot{\gamma}(t_2)}
{2|\gamma(t_1)-\gamma(t_2)|^2}\right) dt_1dt_2=\\
\displaystyle \iint \limits_{e_i \times e_j}
\frac{\dot{\gamma}(t_1)\dot{\gamma}(t_2)}{|\gamma(t_1)-\gamma(t_2)|^2}
dt_1dt_2.
\end{array}
$$

{\bf 3).} Consider now the case of an ordered couple of edges
$e_i$ and $e_j$ adjacent to their common vertex $v$ with the
corresponding angle $\alpha_{ij}$. Let $t_1$ and $t_2$ be some
points of edges $e_i$ and $e_j$ respectively. Orient the edge
$e_i$ in the direction to the common vertex $v$, and the edge
$e_j$ in the direction from the common vertex $v$. Denote by
$C(\gamma(v),\gamma(t_1),\gamma(t_2))$ the circle passing through
the points $\gamma(v)$, $\gamma(t_1)$, and $\gamma(t_2)$ with the
orientation corresponding to the following order of points:
$\gamma(v)$, $\gamma(t_1)$, $\gamma(t_2)$. Denote the angle
between the oriented circles
$C(\gamma(v),\gamma(t_1),\gamma(t_2))$ and
$C(\gamma(t_2),\gamma(t_2),\gamma(t_1))$ by
$\beta_{\gamma,ij}(t_1,t_2)$. M\"obius energy for the ordered
couple of edges $e_i$ and $e_j$ is defined as follows:
$$
\begin{array}{l}
M(\gamma;e_i,e_j)=\\
 \displaystyle \iint \limits_{e_i \times
e_j}\left(
\frac{\dot{\gamma}(t_1)\dot{\gamma}(t_2)}{|\gamma(t_1)-\gamma(t_2)|^2}
-\cos\bigl(\theta_{\gamma}(t_1,t_2){+}2\beta_{\gamma,ij}(t_1,t_2){-}\alpha_{ij}
{-} \pi \bigr)
\frac{\dot{\gamma}(t_1)\dot{\gamma}(t_2)}{|\gamma(t_1)-\gamma(t_2)|^2}\right)
dt_1dt_2.
\end{array}
$$

Again the minuend of the integrand is called the {\it principal
term}, and tee subtrahend is called the {\it normalization term}.

\begin{figure}
$$\epsfbox{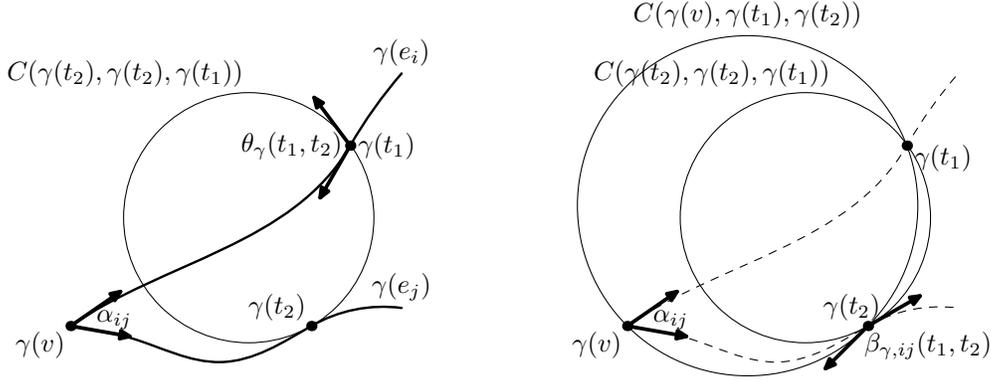}$$
\caption{The angles $\theta_{\gamma}(t_1,t_2)$ and
$\beta_{\gamma,ij}(t_1,t_2)$.}\label{circles}
\end{figure}

\begin{definition}
Let $\gamma : G \to \r ^3$ be a $C^2$-smooth embedding of some
graph $G$. By {\it M\"obius energy} of the of the embedding
$\gamma$ we mean the following value:
$$
M(\gamma,G)=\sum\limits_{i=1}^{n}\sum\limits_{j=1}^{n}M(\gamma;e_i,e_j).
$$
\end{definition}

\section{Main properties of M\"obius energy for graphs.}

Let $G$ be some graph with edges $e_i$ where $i=1, \ldots , n$
and vertices $v_k$ for $k=1, \ldots , m$. We suppose also that the
graph $G$ does not have simple loops and multiple edges. Let
$\gamma:G \longrightarrow \r ^3$ be an $\alpha$-embedding.

\begin{theorem}
The following statements hold.
\\
i$)$. M\"obius energy $M(\gamma,G)$ is well-defined.
\\
ii$)$. The value $M(\gamma,G)$ does not depend on the
parametrization and orientation choice for the edges of the graph
$G$.
\\
iii$)$. M\"obius energy is nonnegative.
\\
iv$)$. Let $T:\r^3\to \r^3$ be some M\"obius transformation such
that the its preimage of infinity does not contain points of the
embedded graph $\gamma(G)$. Then $M(\gamma,G)=M(T\circ\gamma,G)$.
\end{theorem}

\begin{proof} Since for any couple of open edges the integrand is
bounded from above and continuous, all integrals converge (for
adjacent edges the boundedness follows form the explicit angle
calculations for the corresponding embedded angle obtained with
two straight rays). Therefore, M\"obius energy of the graph
$M(\gamma,G)$ is well-defined. This proves the first property.
The second Property is obviously satisfied. Since all integrands
are nonnegative, M\"obius energy is also nonnegative. Hence the
third property holds.

To prove the last property we remind that the set of all straight
lines and circles maps to itself under (conformal) M\"obius
transformations, and also the angles between them are invariant.
Hence the angles $\alpha_{ij}$, $\beta_{\gamma,ij}(t_1,t_2)$ and
$\theta_{\gamma}(t_1,t_2)$ are invariant. Besides that the
cross-ratio is also invariant under M\"obius transformations.
Thus the following value is M\"obius invariant:
$$
\frac{\dot{\gamma}(t_1)\dot{\gamma}(t_2)}{|\gamma(t_1)-\gamma(t_2)|^2}
$$
Therefore, M\"obius energy of graphs is invariant under M\"obius
transformations.
\end{proof}

\begin{remark} If we multiply the integrand by some smooth arbitrary function
of $\theta_{\gamma}$ then we will obtain a new functional on the
space of graphs invariant under M\"obius group action.
\end{remark}

Further we fix all angles at the vertices in families of
$\alpha$-embeddings (i.e. all $\alpha$ are equal to each other).

\begin{theorem} Consider an arbitrary graph $G$ and an arbitrary
set of angles $\alpha$, corresponding to the graph $G$.\\
i$)$. Let $\gamma_{\delta}$ where $0\le \delta\le 1$ be a
continuous family of $\alpha$-embeddings, then the function
$M(\gamma_{\delta},G)$
is continues in the variable $\delta$.\\
ii$)$. Let $\gamma_{\delta}$ for $0\le \delta\le 1$ be a
continuous family of $\alpha$-immersions where $\gamma_0$ is an
immersion with the unique point of transversal double
self-intersection and $\gamma_{\delta}$ for any $\delta \ne 0$ is
an embedding, then
$$\lim\limits_{\delta \to 0}M(\gamma_{\delta},G)=+\infty.$$
\end{theorem}

{\it Sketch of the proof.} {\bf i).} Consider an arbitrary couple
of open edges $e_i$ and~$e_j$. Since the integrand (considered as
a function in $t_1$, $t_2$, and $\delta$ variables) is
continuous, non-negative, and bounded from above on the set
$e_i{\times} e_j {\times} [0,1]$, the first statement of theorem
holds.

{\bf ii).} Let the edges $e_i$ and $e_j$ (where $i \ne j$) of the
immersion $\gamma_0$ transversely intersect with the angle
$\varphi$. Suppose that the edges $e_i$ and $e_j$ are adjacent to
some common edge. Consider an embedding $\gamma_{\delta}$ and
denote the parts of the edges $e_i$ and $e_j$ which images under
$\gamma_{\delta}$ are contained in complement of  the ball of
radius $(\sqrt{\delta}d$) to the ball of radius $d$ both centered
at the self-intersection point of the immersion $\gamma_0$ by
$e_i(d,\delta)$ and $e_{j}(d,\delta)$ respectively.

For any positive $\varepsilon$ there exists some positive $d$ such
that for any positive $\delta<d$ for all points of the set
$e_i(d,\delta){\times} e_{j}(d,\delta)$ holds
$$
\Bigl|\beta_{\gamma_{\delta},ji}(t_1,t_2)-\bigl(\pi{-}\varphi{-}\beta_{\gamma_{\delta},ij}(t_1,t_2)\bigr)\Bigr|
< \varepsilon.
$$
Let us estimate the integral for the corresponding sets
$e_i(d,\delta){\times} e_{j}(d,\delta)$. Let us sum up the
integrands for the ordered couples ($e_i(d,\delta)$,
$e_{j}(d,\delta)$) and ($e_j(d,\delta)$,
$e_{i}(\varepsilon,\delta)$). The normalization term will be as
follows:
$$
\begin{array}{l}
\displaystyle
\bigl(\cos\bigl(\theta_{\gamma_{\delta}}(t_1,t_2){+}2\beta_{\gamma_{\delta},ij}(t_1,t_2){-}\alpha_{ij}
{-} \pi \bigr)+
\cos\bigl(\theta_{\gamma_{\delta}}(t_1,t_2){+}2\beta_{\gamma_{\delta},ji}(t_1,t_2){-}\alpha_{ij}
{-} \pi \bigr)\bigr) \times \\
\displaystyle\frac{\dot{\gamma_{\delta}}(t_1)\dot{\gamma_{\delta}}(t_2)}{|\gamma_{\delta}(t_1)-\gamma_{\delta}(t_2)|^2}.
\end{array}
$$
Since for the set under consideration the following holds:
$$
\Bigl|\beta_{\gamma_{\delta},ji}(t_1,t_2)-\bigl(\pi{-}\varphi{-}\beta_{\gamma_{\delta},ij}(t_1,t_2)\bigr)\Bigr|
< \varepsilon,
$$
the sum of cosines of the normalization term differ from
$$
2\cos\bigl(\theta_{\gamma_{\delta}}(t_1,t_2){-}\alpha_{ij} {-}
\varphi \bigr) \cos\bigl(2\beta_{\gamma_{\delta},ij}{+} \varphi
{-}\pi\bigr)
$$
less than by $\varepsilon$.

The absolute value of the second factor is less than or equal to
the unity. The absolute value of the first factors essentially
varies and there exist a subset of $e_i(d,\delta){\times}
e_{j}(d,\delta)$ of positive measure such that its absolute value
is bounded from above by some constant less than unity. Hence for
any positive $C$ there exist sufficiently small $d$ such that for
$\delta$ tending to zero the value of the functional of M\"obius
energy on the set $e_i(d,\delta'){\times}
e_{j}(\varepsilon,\delta')$ tends to some real number greater
than $C$.  Therefore,
$$\lim\limits_{\delta \to 0}M(\gamma_{\delta},G)=+\infty.$$

The case of self-intersection of one edge is similar to the
classical case of knot M\"obius energy. The case of
self-intersection of two edges that do not have the common vertex
is trivial.

\qed

\section{On some critical objects for M\"obius energy functional.}

In this section we formulate some statements and question related
to the critical configurations of angles at vertices, and discuss
the techniques of construction of special embedded graphs with
critical values of M\"obius energy.

\subsection{Critical configurations of angles at vertices.}

By the {\it intensity} of an angle $\alpha \in (0,\pi]$ we mean
the following value:
$$
\psi (\alpha)= \left\{
\begin{array}{ll}
1-\frac{\pi-\alpha}{\sin (\alpha)}& \mbox{for } 0<\alpha<\pi\\
0& \mbox{for } \alpha=\pi
\end{array}
\right. .
$$

Consider an arbitrary graph $G$ and an embedding $\gamma$. Take a
couple of edges $e_i$ and $e_j$ adjacent to some common vertex
$v$ with the corresponding angle $\alpha_{ij}$. Let
$V_\varepsilon \subset e_i {\times} e_j $ be the set of couples
$(t_1,t_2)$ such that the images $\gamma(t_1)$ and $\gamma(t_2)$
are not contained in the ball of radius $\varepsilon$ centered at
the point $\gamma(v)$.

\begin{statement}\label{zzz}
The integral of the principal term estimates as follows
$$
\iint \limits_{V_\varepsilon}
\frac{\dot{\gamma}(t_1)\dot{\gamma}(t_2)}{|\gamma(t_1)-\gamma(t_2)|^2}
dt_1dt_2= \psi (\alpha (i,j)) \ln \left(
\frac{1}{\varepsilon}\right) + C(\gamma)+ o(1),
$$
while $\varepsilon$ tends to zero. The constant $C(\gamma)$ here
does not depend on $\varepsilon$. \qed
\end{statement}

Now consider the configuration space $\Omega_k$ of $k$-tuples
non-coinciding enumerated unit segments in $\r^3$ with the common
vertex at the origin. Consider an arbitrary $k$-tuple $\omega$ of
$\Omega_k$. Denote the angle between the $i$-th and the $j$-th
segments by $\alpha_{ij}$ where $0<\alpha_{ij}\le \pi$.

\begin{definition}
Let $\omega$ be some $k$-tuple of $\Omega_k$. The following value
$$
\Psi(\omega)=\sum\limits_{i=1}^{k}\sum\limits_{j>i}^{k}\psi(\alpha_{ij})
$$
is called the {\it intensity} of the $k$-tuple $\omega$.
\end{definition}

Let $v$ be some vertex of $G$ of order $k$, and $\gamma$
--- some $\alpha$-embedding of $G$ that maps $v$ to the origin.
Denote by $\omega(\gamma;v)$ the $k$-tuple of $\Omega_k$ whose
segments are tangent to the corresponding edges of $\gamma(G)$ at
the origin.

\begin{definition} A vertex $v$ of the order $k$ is said to be
{\it critical} ({\it extremal}, {\it minimal}), if the $k$-tuple
$\omega(\gamma;v)$ is critical (extremal, minimal) for the
function $\Psi$.
\end{definition}

Since all angles between vectors of the first derivatives are
fixed, the value $\Psi(\omega(\gamma;v))$ does not depend on an
embedding $\gamma$. Therefore, the property of the vertex to be
 critical (extremal, minimal) also does not depend on the embedding $\gamma$.

The question of finding vertices with the least intensity is
natural here. Here we show some examples of extremal vertices.

\begin{statement}
i$)$. The angle at any critical vertex of order two is straight.
This critical vertex is also the minimal vertex.
\\
ii$)$. All angles at any critical vertex of order three are equal
to $\frac{2\pi}{3}$. The critical vertex is also the minimal
vertex at that.
\\
iii$)$. There exist at least two critical configurations of
angles for the vertices of order four:

a$)$ the first one corresponds to the diagonals of the square;

b$)$ the second one corresponds to the segments that join the
mass center of the homogeneous regular tetrahedron with the
vertices of this tetrahedron.  \qed
\end{statement}

The further classification of critical vertices is unknown for the
author.

\begin{problem} Find all angle configurations at vertices
of degree four, five (of degree $n$) that corresponds to critical,
extremal, and minimal vertices. To which convex polyhedra do these
angle configurations correspond?
\end{problem}

It is supposed that the vertex of degree four that corresponds to
the square is not extremal; the vertex of degree four that
corresponds to the regular tetrahedron is minimal; all other
vertices of order four are not critical.

\subsection{Some examples of critical graphs.}

In conclusion of this paper we study some examples of critical
graphs that were constructed by the techniques of D.~Kim and
R.~Kusner~\cite{Kim}. Namely, consider a three-dimensional space
as a subspace a four-dimensional space. By some M\"obius
transformation in the four-dimensional space this subspace is
taken to the unit sphere defined by the following equation:
$x^2{+}y^2{+}z^2{+}t^2{=}1$. This sphere contains a family of
''symmetric tori'' that are obtained as intersections of the
sphere and surfaces of type $x^2{+}y^2{=}\lambda^2$, where
$\lambda\in (0,1)$ is a parameter of the family. The torus
$T_{1/2}$ divides the sphere into two symmetric with respect to
this torus parts. Now by symmetry reasons any symmetric graph of
$T_{1/2}$ that corresponds to some regular quadratic lattice of
$T_{1/2}$ is critical.

The rectangular symmetric graphs of the torus $T_{1/2}$ are
parametrized (up to the length preserving transformations of
$\r^4$) by $4$-tuples of integers $(p,q;m,n)$. Here  $p$ and $q$
are either relatively prime and $p{\ge} q{\ge} 1$ or  $p{=}1$ and
$q{=}0$; and also $m$ and $n$ satisfy the following conditions:
$m{\ge} n{\ge} 1$. Let us call one of the lattice directions of
the rectangular symmetric graph {\it horizontal} and the
perpendicular to it --- {\it vertical}. The couple $(p,q)$ is
defined by the torus winding corresponding to the horizontal
direction. The integers $m$ and $n$ equal to the numbers of
horizontal and respectively vertical ''circles'' in the graph.
Some lattice is square iff $m=n$. So critical graphs are the
graphs of the type $(p,q;n,n)$. This critical graphs are taken to
some critical graph embeddings in $\r^3$ by some stereographic
projection.
\begin{figure}
$$\epsfbox{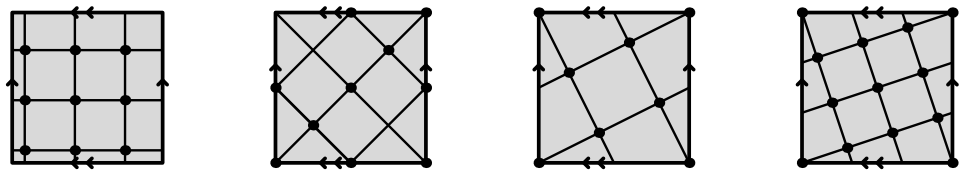}$$
\caption{Symmetric toric graphs $(1,0;3,3)$, $(1,1;2,2)$,
$(2,1;1,1)$ and $(3,1;1,1)$.}\label{tori}
\end{figure}

See the examples of symmetric toric square graphs $(1,0;3,3)$,
$(1,1;2,2)$, $(2,1;1,1)$ and $(3,1;1,1)$  on Fig.~\ref{tori}
(from the left to the right). The approximate calculations show
that
$$
\begin{array}{l}
M((2,1;1,1))\approx 25.137,\\
M((1,1;2,2))\approx 68.789,\\
M((1,0;3,3))\approx 95.979,\\
M((3,1;1,1))\approx 109.91.
\end{array}
$$

\vspace{1cm}

\end{document}